\documentclass[preprint]{ptephy_v2}
\usepackage[colorlinks=true, linkcolor=blue, citecolor=blue, urlcolor=blue]{hyperref}
\usepackage{bm,latexsym,amssymb,amsmath,mathrsfs,mathtools}
\usepackage{graphicx}
\usepackage{color}
\usepackage{orcidlink}  
\usepackage[italicdiff]{physics}

\newcommand{\ri}{\mathrm{i}}

\begin{document}

\title{Anomaly of conserved and nonconserved axial charges in Hamiltonian lattice gauge theory}

\author[1,2]{Yoshimasa Hidaka\orcidlink{0000-0001-7534-6418}}
\author[3]{Arata Yamamoto}

\affil[1]{Yukawa Institute for Theoretical Physics, Kyoto University, Kyoto 606-8502, Japan}
\affil[2]{RIKEN Center for Interdisciplinary Theoretical and Mathematical Sciences (iTHEMS), RIKEN, Wako 351-0198, Japan}
\affil[3]{Department of Physics, The University of Tokyo, Tokyo 113-0033, Japan}

\subjectindex{B01}
\preprintnumber{2506.01336}
%\preprint{RIKEN-iTHEMS-Report-25, YITP-25-84}

\begin{abstract}
We investigate the axial anomaly in Hamiltonian lattice gauge theory. The definition of axial charge operators is ambiguous, especially between conserved and nonconserved axial charges. While these charges appear to differ only by a higher-order term in lattice spacing, they do not coincide in the continuum limit. We demonstrate, through analytical and numerical calculations in $1+1$ dimensions, that the conserved axial charge correctly reproduces the axial anomaly relation in continuous spacetime. Our finding would serve as a valuable lesson about doubler artifact in Hamiltonian time evolution of lattice gauge theory.
\end{abstract}

\maketitle

\section{Introduction}

Traditional lattice gauge theory relies on the path integral formalism and classical numerical techniques, such as Monte Carlo sampling. However, this approach faces computational limitations when addressing specific problems, particularly real-time dynamics.
Hamiltonian lattice gauge theory reformulates lattice gauge theory by quantum states and operators \cite{Kogut:1979wt}.
The Hamiltonian formalism enables the direct application of quantum computing techniques to lattice gauge theory \cite{Banuls:2019bmf,Bauer:2022hpo,DiMeglio:2023nsa}.
The powerful intersection of theoretical physics and quantum technology has the potential to unlock new solutions to longstanding challenges in quantum field theory.

The axial anomaly is one of the most intriguing subjects in lattice gauge theory.
On a lattice, chiral symmetry is not the same as in continuous spacetime but is defined through the Ginsparg-Wilson relation \cite{Ginsparg:1981bj}.
One approach to satisfying the Ginsparg-Wilson relation is the overlap fermion \cite{Neuberger:1997fp,Neuberger:1998wv}, which has been rigorously shown to reproduce the axial anomaly in the path-integral formalism \cite{Luscher:1998pqa,Kikukawa:1998pd,Fujikawa:1998if,Adams:1998eg,Suzuki:1998yz}.
The Hamiltonian formalism of the Ginsparg-Wilson fermion was proposed around the same time \cite{Horvath:1998gq,Cheluvaraja:2000an,Creutz:2001wp}.
Due to the difficulty of full Hamiltonian calculation, an early study derived the axial anomaly only in the classical continuum limit \cite{Matsui:2005uh}.
Recently, the Hamiltonian formalism has seen renewed interest, particularly for the application of quantum computing to chiral lattice fermions \cite{Hayata:2023zuk,Hayata:2023skf,Clancy:2023kla,Singh:2025sye}.

In this paper, we focus on the ambiguity in defining the axial charge operator in the Hamiltonian lattice gauge theory.
In general, lattice operators are not uniquely defined and adding higher-order operators is feasible.
We examine two distinct definitions of axial charges and provide an explicit proof of the axial anomaly relation in the operator formalism.
While these definitions are expected to coincide in the continuum limit, we demonstrate that this is not the case.
We also note that the ambiguity in defining the axial charge has been a topic of interest in other contexts as well \cite{Chatterjee:2024gje,Yamaoka:2025sdm}.

\section{Hamiltonian}

We consider a one-dimensional U$(1)$ lattice gauge theory with a massless fermion.
The lattice spacing is denoted as $a$.
The gauge link operator $U(x)=\exp\{\ri aeA(x)\}$ and its conjugate operator $E(x)$ satisfy the commutation relation 
\begin{equation}
\label{eqcr}
\left[ E(x),U(y) \right]=e\, U(x)\delta_{xy} .
\end{equation}
The gauge coupling constant $e$ has mass dimension one, so $\dim[ea]=0$.
The fermion creation and annihilation operators satisfy the anticommutation relation
\begin{equation}
\label{eqacr}
 \left\{ \psi_\alpha(x) , \psi^\dagger_\beta(y) \right\} =  \frac{1}{a} \delta_{\alpha\beta} \delta_{xy}
\end{equation}
with the two-component spinor indices $\alpha,\beta$.

The total Hamiltonian is given by
\begin{equation}
 H = H_g + H_f
\end{equation}
with the gauge part
\begin{equation}
 H_g = a \sum_{x} \frac12 E(x) E(x)
\end{equation}
and the fermion part
\begin{equation}
\label{eqHf}
 H_f = a \sum_{x,y} \psi^\dagger(x) \gamma^0 D(x,y) \psi(y) .
\end{equation}
The general form of the overlap Dirac operator $D$ is far from simple \cite{Creutz:2001wp}.
In one dimension, however, there is a special property: the overlap fermion can be reduced to simpler lattice fermions.
The massless overlap fermion is identical to the massless Wilson fermion~\cite{Horvath:1998gq} and to the massless staggered fermion in a non-interacting case~\cite{Hayata:2023skf}.
This property significantly simplifies analysis.
We adopt the Wilson-Dirac operator
\begin{equation}
 D = \frac{1}{2} \left\{ \ri\gamma^1 (\nabla+\nabla^\dagger) - a\nabla^\dagger \nabla \right\}
\end{equation}
with the lattice covariant derivative
\begin{eqnarray}
    \nabla \psi(x) &=& \frac{1}{a} \left\{ U(x) \psi(x+a) - \psi(x) \right\}, \\
    \nabla^\dagger \psi(x) &=& \frac{1}{a} \left\{ \psi(x) - U^\dagger(x-a) \psi(x-a) \right\} .
\end{eqnarray}
The massless fermion Hamiltonian \eqref{eqHf} involves only on-site and nearest-neighbor hopping terms:
\begin{equation}
\begin{split}
\label{eqHW}
 H_f &= a\sum_x \bigg\{ \psi^\dagger(x) \gamma^0 \psi(x)
-\frac{1}{2} \psi^\dagger(x) \gamma^0 (1-\ri\gamma^1) U(x) \psi(x+a) \\
&\quad - \frac{1}{2} \psi^\dagger(x+a) \gamma^0 (1+\ri\gamma^1) U^\dagger(x) \psi(x) \bigg\} ,
\end{split}
\end{equation}
where $\gamma^\mu$ are gamma matrices satisfying $(\gamma^0)^2=-(\gamma^1)^2=1$, and $\gamma^0\gamma^1=-\gamma^1\gamma^0$.

The theory possesses the global U$(1)_{\rm V}$ symmetry.
The charge density
\begin{equation}
  \rho(x) = \psi^\dag(x)\psi(x)- \frac{1}{a},
\end{equation}
and the current density
\begin{equation}
\label{eqjv}
\begin{split}
  j(x) &=  \frac{\ri}{2} \psi^\dagger(x) \gamma^0 (1+\ri\gamma^1) U^\dag(x-a) \psi(x-a) \\
  &\quad - \frac{\ri}{2} \psi^\dagger(x-a) \gamma^0(1-\ri\gamma^1) U(x-a) \psi(x) ,
\end{split}
\end{equation}
satisfy the continuity equation
\begin{equation}
\begin{split}
    \partial_t\rho(x) +\partial_xj(x) = \ri [H,\rho(x)] +\partial_xj(x)=0 .
\end{split}
\end{equation}
We define the spatial derivative $\partial_x$ by the forward difference
\begin{equation}
\partial_x j(x)= \frac{1}{a} \{j(x+a)-j(x)\} .
\end{equation}
Summing over the space, we obtain the conserved charge operator
\begin{equation}
 Q = a \sum_{x} \rho(x) ,
\end{equation}
which satisfies the vector charge conservation
\begin{equation}
 \partial_t Q = \ri [H,Q] = 0 .
\end{equation}
As for the global U$(1)_{\rm A}$ symmetry, there is an ambiguity in defining an axial charge operator.
In the path-integral formalism, the conserved axial charge is uniquely defined through symmetry transformation and given by point-splitting in time direction \cite{Kikukawa:1998py}.
The point-split form cannot be straightforwardly converted to an operator in the Hamiltonian formalism, where time remains continuous.
In the Hamiltonian formalism, the conserved charge is defined through its commutation relation with a Hamiltonian.
We introduce two definitions of axial charges: the {\it nonconserved axial charge}
\begin{equation}
  Q_5' = a \sum_{x} \psi^\dagger(x) \gamma^5 \psi(x)
\end{equation}
and {\it conserved axial charge} \cite{Horvath:1998gq,Creutz:2001wp}
\begin{equation}
\label{eqqcon}
 Q_5 = a \sum_{x,y} \psi^\dagger(x) \gamma^5 \left\{ 1-\frac{a}{2}D(x,y) \right\} \psi(y) .
\end{equation}
The former is not conserved,
\begin{equation}
 \partial_t Q_5' = \ri [H_f,Q_5'] \neq 0 ,
\end{equation}
but the latter is conserved,
\begin{equation}
 \partial_t Q_5 = \ri [H_f,Q_5] =0 ,
\end{equation}
for a chiral lattice fermion without gauge interaction.
The two definitions differ by $aD/2$, which is a higher order of $a$.
One might naively expect that they make no difference in the continuum limit, and both reproduce the anomaly when gauge interaction exists.
This is, however, not the case as demonstrated below.

\section{Nonconserved axial charge}

Since the nonconserved axial charge is local and independent of $U(x)$, it commutes with the gauge Hamiltonian $H_g$.
The time evolution of the axial charge density
\begin{equation}
  \rho_5'(x) = \psi^\dagger(x) \gamma^5 \psi(x)
\end{equation}
is governed solely by the fermion Hamiltonian
\begin{equation}
    \partial_t\rho_5'(x)=\ri[H_f,\rho_5'(x)] .
\end{equation}
Let us define the current density as
\begin{align}
\label{eqjnon}
\begin{split}
  j_5'(x) 
  &= - \ri \psi^\dag(x) \gamma^1 \psi(x)
  + \frac12 \psi^\dag(x-a)(1+\ri\gamma^1) U(x-a) \psi(x) \\
  &\quad + \frac12 \psi^\dag(x)(1+\ri\gamma^1) U^\dag(x-a) \psi(x-a) .
\end{split}
\end{align}
This is analogously defined as the vector current density~\eqref{eqjv}, except for the first term.
(The reason for the first term is explained just below.)
The axial current divergence is given as
\begin{equation}
\label{eq:ANC}
\begin{split}
    \partial_\mu j_5'^\mu(x) 
    &\coloneqq \partial_t\rho_5'(x) +\partial_xj_5'(x) \\
    &= \frac{\ri}{a} \{ - \psi^\dag(x)\gamma^1\psi(x)
     - \psi^\dag(x+a)\gamma^1\psi(x+a) \\
    &\quad + \psi^\dag(x) \gamma^1 U(x)\psi(x+a)
     + \psi^\dag(x+a) \gamma^1 U^\dag(x)\psi(x) \}\\
    &= -\ri a ( \nabla \psi(x) )^\dag \gamma^1 \nabla \psi(x) .
\end{split}
\end{equation}
The right-hand side is $O(a)$ because it originates from explicit axial symmetry breaking by the Wilson term.
As shown in Appendix, it vanishes in the continuum limit,
\begin{equation}
    \partial_\mu j_5'^\mu(x) = 0 + O(a).
    \label{eqANC}
\end{equation}
The axial anomaly cannot be obtained.
(If the current density \eqref{eqjnon} does not include the first term, a total spatial divergence remains in Eq.~\eqref{eqANC}.
Such a total divergence can be always absorbed into the definition of the current density.)

Curiously, Eq.~\eqref{eqANC} suggests axial charge conservation in the continuum limit.
Even if an electric field is applied to the vacuum, the axial charge is not produced but conserved.
Does this imply that U(1) gauge theory can be formulated without the mixed anomaly of the U$(1)_{\rm V}$ and U$(1)_{\rm A}$ symmetries?
This puzzle was recognized at an early age of the Hamiltonian lattice gauge theory \cite{Ambjorn:1983hp}.
The Wilson-Dirac Hamiltonian \eqref{eqHW} has doublers with large but finite energy $\varepsilon_d \coloneqq \pm 2/a$.
When the characteristic time scale is shorter than the inverse of the energy gap, $t < 1/|\varepsilon_d|$, the doublers contribute to dynamics and cancel physical contributions, just as naive lattice fermions.
The continuum limit leading to Eq.~\eqref{eqANC} corresponds to such a short time scale.
The authors of Ref.~\cite{Ambjorn:1983hp} proposed that the correct anomaly relation is recovered by time evolution with long time scales s.t.~$t > 1/|\varepsilon_d|$.
In practice, however, tracking long-time evolution is difficult both in analytical and numerical calculations.
We would like to propose a practical and modern solution to resolve this issue.

\section{Conserved axial charge}

From the definition of the conserved axial charge $Q_5$, we introduce the charge density
\begin{equation}
\begin{split}
  \rho_5(x)
  &=\frac12 \psi^\dag(x)\gamma^5 \psi(x)
   + \frac14 \psi^\dag(x)\gamma^5(1-\ri\gamma^1) U(x) \psi(x+a) \\
  &\quad + \frac14 \psi^\dag(x+a)\gamma^5(1+\ri\gamma^1) U^\dag(x) \psi(x),
\end{split}
\end{equation}
and the current density
\begin{equation}
\begin{split}
  j_5(x) 
  &= \frac12 \psi^\dag(x) (1-\ri\gamma^1) \psi(x)
   + \frac14 \psi^\dag(x-a)(1+\ri\gamma^1) U(x-a) \psi(x) \\
  &\quad + \frac14 \psi^\dag(x)(1+\ri\gamma^1) U^\dag(x-a) \psi(x-a) .
\end{split}
\end{equation}
The straightforward calculation shows
\begin{equation}
\label{eqdj}
    \ri[H_f, \rho_5(x)] +\partial_xj_5(x)=0,
\end{equation}
which suggests a continuity equation in a case without gauge interaction.
Since the charge density $\rho_5(x)$ does not commute with either the gauge hamiltonian or the fermion Hamiltonian, its time evolution is given by 
\begin{equation}
\label{eqdrho}
    \partial_t\rho_5(x)=\ri[H_g,\rho_5(x)]+\ri[H_f,\rho_5(x)].
\end{equation}
From Eqs.~\eqref{eqdj} and \eqref{eqdrho}, the axial current divergence can be expressed as
\begin{equation}
    \partial_\mu j_5^\mu(x) 
    \coloneqq \partial_t\rho_5(x) +\partial_xj_5(x) 
    =\ri [ H_g,\rho_5(x)].
\end{equation}
Applying the canonical commutation relation \eqref{eqcr}, we obtain
\begin{equation}
\label{eq:AC}
\begin{split}
    \partial_\mu j_5^\mu(x) 
    &= \frac{\ri ae}{8} \big[ \psi^\dag(x)\gamma^5(1-\ri\gamma^1)\{E(x)U(x) +U(x)E(x)\}\psi(x+a) \\
    &\quad -\psi^\dag(x+a)\gamma^5(1+\ri\gamma^1) \{E(x)U^\dagger(x)+U^\dagger(x)E(x)\}\psi(x)
  \big].
\end{split}
\end{equation}
Although the right-hand side appears to be proportional to $a$, it does not vanish in the continuum limit $a\to 0$.
To see this, we decompose a fermion bilinear into its normal ordered and divergent components as
\begin{equation}
  \psi_\alpha^\dag(x)\psi_\beta(y)
  =:\psi_\alpha^\dag(x)\psi_\beta(y):
  +\delta [ \psi_\alpha^\dag(x)\psi_\beta(y) ].
\end{equation}
As derived in Appendix, the fermion bilinears in Eq.~\eqref{eq:AC} have $O(1/a)$ divergence
\begin{equation}
\label{eq:delta}
    \delta [ \ri \psi^\dag(x)\gamma^5(1-\ri\gamma^1)\psi(x+a) ] 
    =\delta [-\ri \psi^\dag(x+a)\gamma^5(1+\ri\gamma^1)\psi(x) ]
    = \frac{2}{a\pi}.
\end{equation}
Therefore, Eq.~\eqref{eq:AC} becomes
\begin{equation}
  \partial_\mu j_5^\mu(x) = \frac{e}{4\pi} \{E(x)U(x)+U(x)E(x) +E(x)U^\dagger(x)+U^\dagger(x)E(x)\} + O(a) .
\end{equation}
In the continuum limit, we obtain the correct anomaly relation
\begin{equation}
\label{eqAC}
  \partial_\mu j_5^\mu(x) = \frac{e}{\pi}E(x) + O(a).
\end{equation}

\section{Classical background}

To understand more about the distinction between the two axial charges, we numerically solved the time evolution of the Wilson-Dirac Hamiltonian \eqref{eqHW} with a classical gauge field.
The classical link variable is set to
\begin{equation}
\label{eqclassical}
    U(x)=
    \begin{cases}
    0 & (t < 0), \\
    \exp(\ri aeEt) & (t \ge 0).
    \end{cases}
\end{equation}
with a uniform electric field $E$.
With this form of the link variable, single-particle states can be obtained by diagonalizing the Hamiltonian using the Fourier transformation.
The initial condition is the vacuum with $\langle Q \rangle=\langle Q_5 \rangle=\langle Q_5' \rangle=0$, where negative-energy states are fully occupied and zero-energy states are half occupied.
The lattice volume is $V=Na=100a$ and the boundary condition is periodic.
We set $a^2eE=0.1$, so the expected anomalous production rate is approximately $a^2eEN/\pi = 10/\pi \simeq 3.18$ in the continuum limit.

\begin{figure}[ht]
\begin{minipage}{0.5\textwidth}
\begin{center}
\includegraphics[width=1\textwidth]{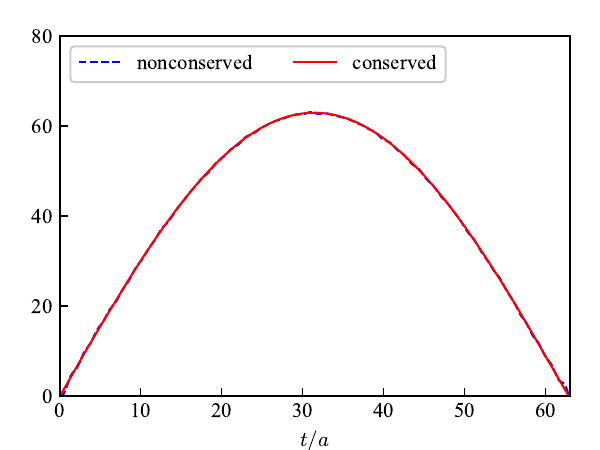}
\end{center}
\end{minipage}
%\end{figure}
%\begin{figure}[ht]
\begin{minipage}{0.5\textwidth}
\begin{center}
\includegraphics[width=1\textwidth]{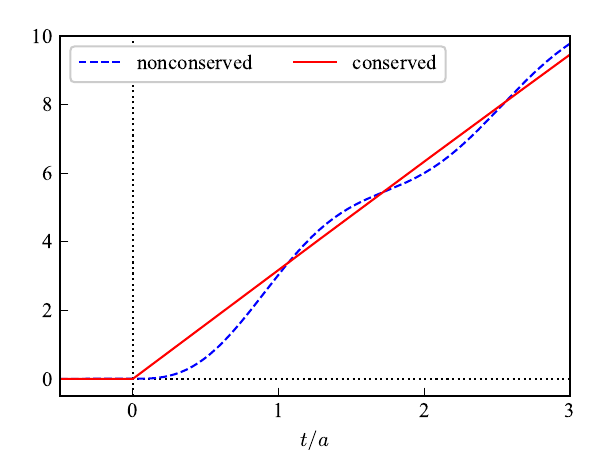}
\end{center}
\end{minipage}
\caption{
Left: anomalous production of the nonconserved axial charge $\langle Q_5'\rangle a$ and the conserved axial charge $\langle Q_5 \rangle a$ against time $t$.
An electric field is switched on at $t=0$.
Right: same as the left panel but the initial region is enlarged.
}
\label{figclassical}
\end{figure}

Figure \ref{figclassical} presents the time evolution of the axial charges.
The axial charges are produced by the electric field in $t>0$.
Since lattice momentum is periodic, the time evolution is given by a periodic function with a period of $t=4\pi/aeE$.
The figure displays a half period of the cycle.
The time region of $t \ll \pi/aeE$ must be used in practice to avoid the lattice artifact.
Although the global behavior is common in the left panel of Fig.~\ref{figclassical}, a closer look reveals differences between the two charges, as seen in the right panel.

As for the nonconserved axial charge, the anomalous production rate at $t=0$ is zero,
\begin{equation}
    \partial_t \langle Q'_5\rangle|_{t=0}=0 ,
\end{equation}
as being consistent with the analytical prediction discussed above.
This consistency can be understood from the explicit form of Eq.~\eqref{eqclassical}, where the limits of $a\to 0$ and $t\to 0$ are equivalent.
In the region of $a/2 < t \ll \pi/aeE$, the production rate approximately takes the expected value $\partial_t \langle Q'_5\rangle \simeq 10/\pi$, but the actual time evolution is contaminated by a small oscillation.
Such oscillatory behavior is a typical feature of the nonconserved axial charge \cite{Hayata:2023skf}.
To extract the correct production rate, it is necessary to simulate long time evolution and to average over a long time period.
This would make the nonconserved axial charge impractical for numerical studies.

\begin{figure}[ht]
\begin{center}
\includegraphics[width=0.6\textwidth]{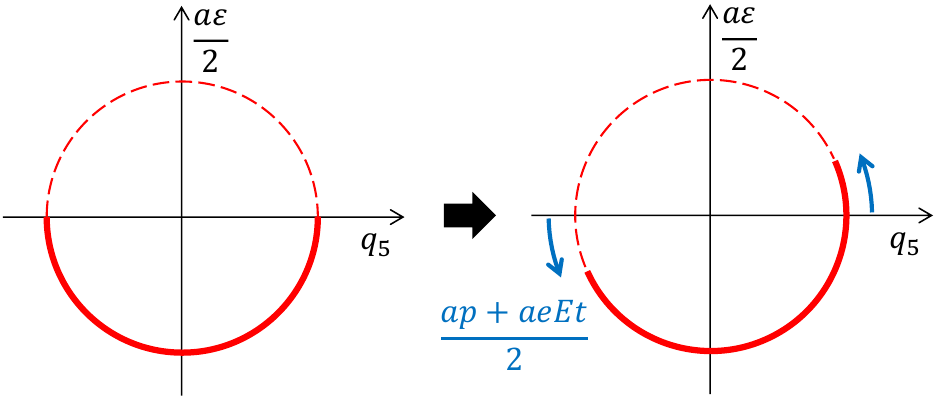}
\end{center}
\caption{
Single-particle spectrum of the energy $\varepsilon$ vs the axial charge $q_5$.
The left is the vacuum state ($t=0$) and the right is an evolved state ($t>0$).
}
\label{figspectrum}
\end{figure}

On the other hand, the production rate of the conserved axial charge immediately aligns with the expected value at $t=0$.
The reason can be simply explained as follows.
Since the Hamiltonian and the conserved axial charge commute, $[H_f,Q_5]=0$, the Hilbert space is described by their simultaneous eigenstates.
In $t\ge 0$, the single-particle energy and axial charge are written as
\begin{align}
\label{eqsinglee}
\varepsilon(p) &=\frac{2}{a} \sin\left(\frac{ap+aeEt}{2}\right) ,
\\
\label{eqsingleq}
q_5(p) &=\cos\left(\frac{ap+aeEt}{2}\right) ,
\end{align}
respectively, with $-2\pi/a \le p < 2\pi/a$.
The single-particle spectrum forms the unit circle $(a\varepsilon/2)^2+q_5^2=1$ \cite{Creutz:2001wp} depicted in Fig.~\ref{figspectrum}.
Initially, at $t=0$, negative-energy states are occupied and the total axial charge is zero (left panel).
After time evolution, the spectrum flows with momentum shift $p \to p+eEt$, leading to a nonzero axial charge (right panel).
The production rate is given by the elementary integral
\begin{equation}
    \partial_t \langle Q_5 \rangle |_{t=0} = \frac{V}{2\pi} \int_{-\frac{2\pi}{a}}^{0} dp  \partial_t q_5(p) |_{t=0} = \frac{eEV}{\pi} ,
\end{equation}
which precisely accounts for the anomaly induced by the classical background \cite{Matsui:2005uh}.
As clearly seen from Fig.~\ref{figspectrum}, only the physical zero modes ($\varepsilon=0$ and $q_5=\pm 1$) contribute to the axial charge production and the doublers ($\varepsilon=\pm 2/a$ and $q_5=\pm 0$) are irrelevant.

We have explicitly demonstrated with a $(1+1)$-dimensional Dirac fermion because of its simplicity.
The above scenario would hold true in $3+1$ dimensions.
Although the explicit forms of the Hamiltonian and the conserved axial charge are complicated, numerical simulations with classical gauge fields are feasible.
The spectral flow under perpendicular electric and magnetic fields has been shown to exhibit the expected behavior \cite{Hayata:2023zuk}.
This reinforces the expectation that the overlap fermion and its conserved axial charge reproduce the axial anomaly in $3+1$ dimensions.

\section{Summary and lesson}

We have demonstrated that the axial anomaly presents a troublesome issue in Hamiltonian lattice gauge theory and that the issue can be resolved by using the conserved axial charge of chiral lattice fermions.
This finding is particularly valuable for the application to quantum computing, of which simulation time is severely limited.
Moreover, it provides a broader lesson for Hamiltonian lattice gauge theory.
Since time direction is not discretized but remains continuous in the Hamiltonian formalism, one might expect that there would be no concern about doubler artifacts arising from time direction.
The axial anomaly serves as a counterexample, implying that this expectation is too naive.
Doublers influence time evolution even for the continuous time-dependent potential $U(x)=\exp(\ri aeEt)$.
If one aims to study a Hamiltonian with discontinuous time dependence, such as quench dynamics, the doubler artifact may be more dangerous.
It is crucial to verify carefully that the artifact vanishes in the continuum limit.

\ack
We would like to express our gratitude to Tomoya Hayata for inspiring this research.
This work was supported by JSPS KAKENHI Grant No.~24H00975 and No.~25K07295, and by JST CREST Grant No.~JPMJCR24I3, Japan.
The work of YH was also partially supported by Center for Gravitational Physics and Quantum Information (CGPQI) at Yukawa Institute for Theoretical Physics.

\appendix
\section{Calculation of fermion bilinear operators}

A fermion bilinear operator is generally decomposed as
\begin{equation}
  \psi_\alpha^\dag(x)\psi_\beta(y)
  =:\psi_\alpha^\dag(x)\psi_\beta(y):
  +\delta [ \psi_\alpha^\dag(x)\psi_\beta(y) ].
\end{equation}
The first term is regular and the second term is singular in $a\to 0$.
Below, we evaluate the singular components in Eqs.~\eqref{eq:ANC} and \eqref{eq:AC}.

For clarity, we first consider a finite lattice volume $V=Na$ with a periodic boundary condition, and take the infinite-volume limit $N\to \infty$ at the end of the calculation.
We adopt the gamma matrix convention $\gamma^0=\sigma_1$, $\gamma^1=-\ri\sigma_3$, and $\gamma^5=\sigma_2$.
Since the link variable is expanded as $U(x)=\exp\{\ri aeA(x)\} =1+O(a)$, we can approximate $U(x)=1$ for leading-order evaluation.
The free Wilson-Dirac equation in momentum space,
\begin{equation}
    \begin{pmatrix}
        -a\varepsilon(p) & 1-e^{\ri ap}\\
        1-e^{-\ri ap} & -a\varepsilon(p)
    \end{pmatrix}
    \begin{pmatrix}
        \psi_1(p)\\
        \psi_2(p)
    \end{pmatrix}
    =0,
\end{equation}
has the energy eigenvalues
\begin{equation}
\varepsilon(p) = \pm \frac{2}{a} \sin\frac{ap}{2},
\end{equation}
and the eigenvectors
\begin{equation}
  \tilde{u}_R(p)=\frac{1}{\sqrt{2}}\begin{pmatrix}
        1\\
        \ri e^{-\ri\frac{ap}{2}}
    \end{pmatrix},\quad
    \tilde{u}_L(p)=\frac{1}{\sqrt{2}}\begin{pmatrix}
      1\\
      -\ri e^{-\ri\frac{ap}{2}}
  \end{pmatrix}.
\end{equation}
The momentum is quantized as $p=2\pi n/V$ with $n=0,\pm 1,\pm 2, \pm (N-1)/2$ due to the periodic boundary condition. 
In this appendix, we assume that $N$ is odd for simplicity.
(Note that the notation is different from Eqs.~\eqref{eqsinglee} and \eqref{eqsingleq}, where $-2\pi\le p<2\pi$, whereas $-\pi<p<\pi$ here.)

The fermion field operator $\psi(x)$ is expanded as
\begin{equation}
\psi(x) = \frac{1}{\sqrt{V}}\sum_{p}\qty(
  \tilde{a}_R(p)\tilde{u}_R(p)e^{ \ri px}+\tilde{a}_L(p)\tilde{u}_L(p)e^{\ri px}
),
\label{eq:wavefunction}
\end{equation}
where $\tilde{a}^\dag_{R/L}(p)$ and $\tilde{a}_{R/L}(p)$ are creation and annihilation operators satisfying the anticommutation relation,
\begin{equation}
    \{\tilde{a}_R(p),\tilde{a}^\dag_R(p')\}=\delta_{pp'},\quad \{\tilde{a}_L(p),\tilde{a}^\dag_L(p')\}=\delta_{pp'},
\end{equation}
and others are zero.
From these commutation relations, we recover Eq.~\eqref{eqacr}.
Each component of $\psi(x)$ is expressed as
\begin{equation}
  \begin{split}
    \psi_1(x)&=\frac{1}{\sqrt{V}}\sum_p\frac{1}{\sqrt{2}}\qty(\tilde{a}_R(p)+\tilde{a}_L(p))e^{\ri px},\\
    \psi_2(x)&=\frac{1}{\sqrt{V}}\sum_p\frac{1}{\sqrt{2}}\qty( \tilde{a}_R(p)- \tilde{a}_L(p))\ri e^{-\ri\frac{ap}{2}}e^{\ri px}.
  \end{split}
\end{equation}
We note that this convention is different from the relativistic one.
We use the nonrelativistic normalization for the wave function, i.e.,
$\tilde{u}^\dag(p)\tilde{u}(p)=1$, while the relativistic normalization for the wave function $u(p)$ is $u^\dag(p)u(p)=2|a\varepsilon(p)|$.
We further decompose into the positive and negative energy contributions,
\begin{equation}
    \tilde{a}_{R/L}(p)=a_{R/L}(p)+b^\dag_{R/L}(-p)
\end{equation}
with
\begin{equation}\label{eq:decomposition_a}
  \begin{split}
    a_R(p) &= \theta(p)\tilde{a}_R(p),\quad a_L(p) = \theta(-p)\tilde{a}_L(p),\\
    b^\dag_R(-p) &= \theta(-p)\tilde{a}_R(p),\quad b^\dag_L(p) = \theta(p)\tilde{a}_L(p),    
  \end{split}
\end{equation}
where $\theta(p)$ is the step function,
\begin{equation}
    \theta(p)=\begin{cases}
        1 & p\geq0\\
        0 & p<0
    \end{cases}.
    \label{eq:stepfunction}
\end{equation}
The ground state $\ket{\Omega}$ satisfies $a_{R/L}(p)\ket{\Omega}=0$ and $b_{R/L}(p)\ket{\Omega}=0$ for all $p$.

In Eq.~\eqref{eq:ANC}, there are four terms to be evaluated.
The first term is explicitly computed as
\begin{equation}
  \begin{split}
    \ri \psi^\dag(x)\gamma^1\psi(x) 
    &=\psi^\dag_1(x)\psi_1(x)-\psi^\dag_2(x)\psi_2(x) \\
    &=\frac{1}{V}\sum_{p,p'}
    (\tilde{a}^\dag_R(p')\tilde{a}_L(p)+\tilde{a}^\dag_L(p')\tilde{a}_R(p)) e^{\ri (p-p')x} \\
    &=:\psi^\dag_1(x)\psi_1(x):-:\psi^\dag_2(x)\psi_2(x): = :\ri \psi^\dag(x)\gamma^1\psi(x) : ,
  \end{split}
\end{equation}
and thus does not have a singular contribution.
The other terms are also regular,
\begin{align}
    \ri \psi^\dag(x+a)\gamma^1\psi(x+a) &= :\ri \psi^\dag(x+a)\gamma^1\psi(x+a): \, ,\\
    \ri \psi^\dag(x+a)\gamma^1\psi(x) &= :\ri \psi^\dag(x+a)\gamma^1\psi(x): \, ,\\
    \ri \psi^\dag(x)\gamma^1\psi(x+a) &= :\ri \psi^\dag(x)\gamma^1\psi(x+a): .
\end{align}
Summarizing these results and using $U(x) =1+O(a)$, we obtain
\begin{equation}
\begin{split}
    -\ri ( \nabla \psi(x) )^\dag \gamma^1 \nabla \psi(x)
    &= - \frac{\ri}{a^2} :\{ \psi^\dag(x)\gamma^1\psi(x)  + \psi^\dag(x+a)\gamma^1\psi(x+a) \\
    &\quad - \psi^\dag(x) \gamma^1 \psi(x+a) - \psi^\dag(x+a) \gamma^1 \psi(x) \}: + O(a) \\
    &= -\ri : ( \partial_x \psi(x) )^\dag \gamma^1 \partial_x \psi(x) : + O(a) .
\end{split}
\end{equation}
Therefore, the right-hand side of Eq.~\eqref{eq:ANC} is $O(a)$ and goes to zero in the continuum limit.

For Eq.~\eqref{eq:AC}, we need to evaluate
\begin{equation}
\begin{split}
    -\ri \psi^\dag(x+a)\gamma^5(1+\ri\gamma^1)\psi(x)
    &=2\psi_2^\dag(x+a)\psi_1(x) \\
    &=\frac{1}{V}\sum_{p,p'} -\ri \Bigl(
\tilde{a}^\dag_R(p)\tilde{a}_R(p')+\tilde{a}^\dag_R(p)\tilde{a}_L(p') \\
&\quad - \tilde{a}^\dag_L(p)\tilde{a}_R(p')- \tilde{a}^\dag_L(p)\tilde{a}_L(p') \Bigr) e^{-\ri\frac{ap}{2}} e^{\ri (p'-p)x},
\end{split}
\end{equation}
and its hermitian conjugate.
Using
\begin{equation}
  \begin{split}
    \tilde{a}^\dag_{R}(p)\tilde{a}_R(p')
    &=a^\dag_{R}(p)a_{R}(p')+a^\dag_{R}(p)b^\dag_{R}(-p')  +b_{R}(-p)a_{R}(p')+b_{R}(-p)b^\dag_{R}(-p')\\
    &=:\tilde{a}^\dag_{R}(p)\tilde{a}_R(p'): + \{b_{R}(-p),b^\dag_{R}(-p')\}\\
    &=:\tilde{a}^\dag_{R}(p)\tilde{a}_R(p'): +\theta(-p)\delta_{p,p'},
  \end{split}
\end{equation}
and
\begin{equation}
  \begin{split}
    \tilde{a}^\dag_{L}(p)\tilde{a}_L(p')
    &=a^\dag_{L}(p)a_{L}(p')+a^\dag_{L}(p)b^\dag_{L}(-p') +b_{L}(-p)a_{L}(p')+b_{L}(-p)b^\dag_{L}(-p')\\
    &=:\tilde{a}^\dag_{L}(p)\tilde{a}_L(p'): + \{b_{L}(-p),b^\dag_{L}(-p')\}\\
    &=:\tilde{a}^\dag_{L}(p)\tilde{a}_L(p'): +\theta(p)\delta_{p,p'},
  \end{split}
\end{equation}
we find the divergent contribution
\begin{equation}
  \begin{split}
     \delta [ -\ri \psi^\dag(x+a)\gamma^5(1+\ri\gamma^1)\psi(x) ]
    &=\frac{1}{V}\sum_{p,p'} -\ri (  \theta(-p)\delta_{p,p'}- \theta(p)\delta_{p,p'} ) e^{-\ri\frac{ap}{2}} e^{\ri (p'-p)x}\\
    &=\frac{2}{V}\sum_{p>0}\sin\frac{ap}{2} \\
    &=\frac{2}{a} \frac{1}{2N\tan\frac{\pi}{2N}}.
  \end{split}
\end{equation}
The same is true for the hermitian conjugate.
Taking the limit of $N \to \infty$, we finally obtain 
\begin{equation}
    \delta [ -\ri \psi^\dag(x+a)\gamma^5(1+\ri\gamma^1)\psi(x) ] = \delta [ \ri \psi^\dag(x)\gamma^5(1-\ri\gamma^1)\psi(x+a) ] = \frac{2}{a\pi} .
\end{equation}

\bibliographystyle{utphys}
\bibliography{paper}

%apsrev4-2.bst 2019-01-14 (MD) hand-edited version of apsrev4-1.bst
%Control: key (0)
%Control: author (72) initials jnrlst
%Control: editor formatted (1) identically to author
%Control: production of article title (-1) disabled
%Control: page (0) single
%Control: year (1) truncated
%Control: production of eprint (0) enabled
\begin{thebibliography}{24}%
\makeatletter
\providecommand \@ifxundefined [1]{%
 \@ifx{#1\undefined}
}%
\providecommand \@ifnum [1]{%
 \ifnum #1\expandafter \@firstoftwo
 \else \expandafter \@secondoftwo
 \fi
}%
\providecommand \@ifx [1]{%
 \ifx #1\expandafter \@firstoftwo
 \else \expandafter \@secondoftwo
 \fi
}%
\providecommand \natexlab [1]{#1}%
\providecommand \enquote  [1]{``#1''}%
\providecommand \bibnamefont  [1]{#1}%
\providecommand \bibfnamefont [1]{#1}%
\providecommand \citenamefont [1]{#1}%
\providecommand \href@noop [0]{\@secondoftwo}%
\providecommand \href [0]{\begingroup \@sanitize@url \@href}%
\providecommand \@href[1]{\@@startlink{#1}\@@href}%
\providecommand \@@href[1]{\endgroup#1\@@endlink}%
\providecommand \@sanitize@url [0]{\catcode `\\12\catcode `\$12\catcode `\&12\catcode `\#12\catcode `\^12\catcode `\_12\catcode `\%12\relax}%
\providecommand \@@startlink[1]{}%
\providecommand \@@endlink[0]{}%
\providecommand \url  [0]{\begingroup\@sanitize@url \@url }%
\providecommand \@url [1]{\endgroup\@href {#1}{\urlprefix }}%
\providecommand \urlprefix  [0]{URL }%
\providecommand \Eprint [0]{\href }%
\providecommand \doibase [0]{https://doi.org/}%
\providecommand \selectlanguage [0]{\@gobble}%
\providecommand \bibinfo  [0]{\@secondoftwo}%
\providecommand \bibfield  [0]{\@secondoftwo}%
\providecommand \translation [1]{[#1]}%
\providecommand \BibitemOpen [0]{}%
\providecommand \bibitemStop [0]{}%
\providecommand \bibitemNoStop [0]{.\EOS\space}%
\providecommand \EOS [0]{\spacefactor3000\relax}%
\providecommand \BibitemShut  [1]{\csname bibitem#1\endcsname}%
\let\auto@bib@innerbib\@empty
%</preamble>
\bibitem [{\citenamefont {Kogut}(1979)}]{Kogut:1979wt}%
  \BibitemOpen
  \bibfield  {author} {\bibinfo {author} {\bibfnamefont {J.~B.}\ \bibnamefont {Kogut}},\ }\href {https://doi.org/10.1103/RevModPhys.51.659} {\bibfield  {journal} {\bibinfo  {journal} {Rev. Mod. Phys.}\ }\textbf {\bibinfo {volume} {51}},\ \bibinfo {pages} {659} (\bibinfo {year} {1979})}\BibitemShut {NoStop}%
\bibitem [{\citenamefont {Ba\~nuls}\ \emph {et~al.}(2020)\citenamefont {Ba\~nuls} \emph {et~al.}}]{Banuls:2019bmf}%
  \BibitemOpen
  \bibfield  {author} {\bibinfo {author} {\bibfnamefont {M.~C.}\ \bibnamefont {Ba\~nuls}} \emph {et~al.},\ }\href {https://doi.org/10.1140/epjd/e2020-100571-8} {\bibfield  {journal} {\bibinfo  {journal} {Eur. Phys. J. D}\ }\textbf {\bibinfo {volume} {74}},\ \bibinfo {pages} {165} (\bibinfo {year} {2020})},\ \Eprint {https://arxiv.org/abs/1911.00003} {arXiv:1911.00003 [quant-ph]} \BibitemShut {NoStop}%
\bibitem [{\citenamefont {Bauer}\ \emph {et~al.}(2023)\citenamefont {Bauer} \emph {et~al.}}]{Bauer:2022hpo}%
  \BibitemOpen
  \bibfield  {author} {\bibinfo {author} {\bibfnamefont {C.~W.}\ \bibnamefont {Bauer}} \emph {et~al.},\ }\href {https://doi.org/10.1103/PRXQuantum.4.027001} {\bibfield  {journal} {\bibinfo  {journal} {PRX Quantum}\ }\textbf {\bibinfo {volume} {4}},\ \bibinfo {pages} {027001} (\bibinfo {year} {2023})},\ \Eprint {https://arxiv.org/abs/2204.03381} {arXiv:2204.03381 [quant-ph]} \BibitemShut {NoStop}%
\bibitem [{\citenamefont {Di~Meglio}\ \emph {et~al.}(2024)\citenamefont {Di~Meglio} \emph {et~al.}}]{DiMeglio:2023nsa}%
  \BibitemOpen
  \bibfield  {author} {\bibinfo {author} {\bibfnamefont {A.}~\bibnamefont {Di~Meglio}} \emph {et~al.},\ }\href {https://doi.org/10.1103/PRXQuantum.5.037001} {\bibfield  {journal} {\bibinfo  {journal} {PRX Quantum}\ }\textbf {\bibinfo {volume} {5}},\ \bibinfo {pages} {037001} (\bibinfo {year} {2024})},\ \Eprint {https://arxiv.org/abs/2307.03236} {arXiv:2307.03236 [quant-ph]} \BibitemShut {NoStop}%
\bibitem [{\citenamefont {Ginsparg}\ and\ \citenamefont {Wilson}(1982)}]{Ginsparg:1981bj}%
  \BibitemOpen
  \bibfield  {author} {\bibinfo {author} {\bibfnamefont {P.~H.}\ \bibnamefont {Ginsparg}}\ and\ \bibinfo {author} {\bibfnamefont {K.~G.}\ \bibnamefont {Wilson}},\ }\href {https://doi.org/10.1103/PhysRevD.25.2649} {\bibfield  {journal} {\bibinfo  {journal} {Phys. Rev. D}\ }\textbf {\bibinfo {volume} {25}},\ \bibinfo {pages} {2649} (\bibinfo {year} {1982})}\BibitemShut {NoStop}%
\bibitem [{\citenamefont {Neuberger}(1998{\natexlab{a}})}]{Neuberger:1997fp}%
  \BibitemOpen
  \bibfield  {author} {\bibinfo {author} {\bibfnamefont {H.}~\bibnamefont {Neuberger}},\ }\href {https://doi.org/10.1016/S0370-2693(97)01368-3} {\bibfield  {journal} {\bibinfo  {journal} {Phys. Lett. B}\ }\textbf {\bibinfo {volume} {417}},\ \bibinfo {pages} {141} (\bibinfo {year} {1998}{\natexlab{a}})},\ \Eprint {https://arxiv.org/abs/hep-lat/9707022} {arXiv:hep-lat/9707022} \BibitemShut {NoStop}%
\bibitem [{\citenamefont {Neuberger}(1998{\natexlab{b}})}]{Neuberger:1998wv}%
  \BibitemOpen
  \bibfield  {author} {\bibinfo {author} {\bibfnamefont {H.}~\bibnamefont {Neuberger}},\ }\href {https://doi.org/10.1016/S0370-2693(98)00355-4} {\bibfield  {journal} {\bibinfo  {journal} {Phys. Lett. B}\ }\textbf {\bibinfo {volume} {427}},\ \bibinfo {pages} {353} (\bibinfo {year} {1998}{\natexlab{b}})},\ \Eprint {https://arxiv.org/abs/hep-lat/9801031} {arXiv:hep-lat/9801031} \BibitemShut {NoStop}%
\bibitem [{\citenamefont {Luscher}(1998)}]{Luscher:1998pqa}%
  \BibitemOpen
  \bibfield  {author} {\bibinfo {author} {\bibfnamefont {M.}~\bibnamefont {Luscher}},\ }\href {https://doi.org/10.1016/S0370-2693(98)00423-7} {\bibfield  {journal} {\bibinfo  {journal} {Phys. Lett. B}\ }\textbf {\bibinfo {volume} {428}},\ \bibinfo {pages} {342} (\bibinfo {year} {1998})},\ \Eprint {https://arxiv.org/abs/hep-lat/9802011} {arXiv:hep-lat/9802011} \BibitemShut {NoStop}%
\bibitem [{\citenamefont {Kikukawa}\ and\ \citenamefont {Yamada}(1999{\natexlab{a}})}]{Kikukawa:1998pd}%
  \BibitemOpen
  \bibfield  {author} {\bibinfo {author} {\bibfnamefont {Y.}~\bibnamefont {Kikukawa}}\ and\ \bibinfo {author} {\bibfnamefont {A.}~\bibnamefont {Yamada}},\ }\href {https://doi.org/10.1016/S0370-2693(99)00021-0} {\bibfield  {journal} {\bibinfo  {journal} {Phys. Lett. B}\ }\textbf {\bibinfo {volume} {448}},\ \bibinfo {pages} {265} (\bibinfo {year} {1999}{\natexlab{a}})},\ \Eprint {https://arxiv.org/abs/hep-lat/9806013} {arXiv:hep-lat/9806013} \BibitemShut {NoStop}%
\bibitem [{\citenamefont {Fujikawa}(1999)}]{Fujikawa:1998if}%
  \BibitemOpen
  \bibfield  {author} {\bibinfo {author} {\bibfnamefont {K.}~\bibnamefont {Fujikawa}},\ }\href {https://doi.org/10.1016/S0550-3213(99)00042-5} {\bibfield  {journal} {\bibinfo  {journal} {Nucl. Phys. B}\ }\textbf {\bibinfo {volume} {546}},\ \bibinfo {pages} {480} (\bibinfo {year} {1999})},\ \Eprint {https://arxiv.org/abs/hep-th/9811235} {arXiv:hep-th/9811235} \BibitemShut {NoStop}%
\bibitem [{\citenamefont {Adams}(2002)}]{Adams:1998eg}%
  \BibitemOpen
  \bibfield  {author} {\bibinfo {author} {\bibfnamefont {D.~H.}\ \bibnamefont {Adams}},\ }\href {https://doi.org/10.1006/aphy.2001.6209} {\bibfield  {journal} {\bibinfo  {journal} {Annals Phys.}\ }\textbf {\bibinfo {volume} {296}},\ \bibinfo {pages} {131} (\bibinfo {year} {2002})},\ \Eprint {https://arxiv.org/abs/hep-lat/9812003} {arXiv:hep-lat/9812003} \BibitemShut {NoStop}%
\bibitem [{\citenamefont {Suzuki}(1999)}]{Suzuki:1998yz}%
  \BibitemOpen
  \bibfield  {author} {\bibinfo {author} {\bibfnamefont {H.}~\bibnamefont {Suzuki}},\ }\href {https://doi.org/10.1143/PTP.102.141} {\bibfield  {journal} {\bibinfo  {journal} {Prog. Theor. Phys.}\ }\textbf {\bibinfo {volume} {102}},\ \bibinfo {pages} {141} (\bibinfo {year} {1999})},\ \Eprint {https://arxiv.org/abs/hep-th/9812019} {arXiv:hep-th/9812019} \BibitemShut {NoStop}%
\bibitem [{\citenamefont {Horvath}\ and\ \citenamefont {Thacker}(1999)}]{Horvath:1998gq}%
  \BibitemOpen
  \bibfield  {author} {\bibinfo {author} {\bibfnamefont {I.}~\bibnamefont {Horvath}}\ and\ \bibinfo {author} {\bibfnamefont {H.~B.}\ \bibnamefont {Thacker}},\ }\href {https://doi.org/10.1016/S0920-5632(99)85172-X} {\bibfield  {journal} {\bibinfo  {journal} {Nucl. Phys. B Proc. Suppl.}\ }\textbf {\bibinfo {volume} {73}},\ \bibinfo {pages} {682} (\bibinfo {year} {1999})},\ \Eprint {https://arxiv.org/abs/hep-lat/9809108} {arXiv:hep-lat/9809108} \BibitemShut {NoStop}%
\bibitem [{\citenamefont {Cheluvaraja}\ and\ \citenamefont {Hari~Dass}(2001)}]{Cheluvaraja:2000an}%
  \BibitemOpen
  \bibfield  {author} {\bibinfo {author} {\bibfnamefont {S.}~\bibnamefont {Cheluvaraja}}\ and\ \bibinfo {author} {\bibfnamefont {N.~D.}\ \bibnamefont {Hari~Dass}},\ }\href {https://doi.org/10.1016/S0550-3213(00)00772-0} {\bibfield  {journal} {\bibinfo  {journal} {Nucl. Phys. B}\ }\textbf {\bibinfo {volume} {598}},\ \bibinfo {pages} {134} (\bibinfo {year} {2001})},\ \Eprint {https://arxiv.org/abs/hep-lat/0001016} {arXiv:hep-lat/0001016} \BibitemShut {NoStop}%
\bibitem [{\citenamefont {Creutz}\ \emph {et~al.}(2002)\citenamefont {Creutz}, \citenamefont {Horvath},\ and\ \citenamefont {Neuberger}}]{Creutz:2001wp}%
  \BibitemOpen
  \bibfield  {author} {\bibinfo {author} {\bibfnamefont {M.}~\bibnamefont {Creutz}}, \bibinfo {author} {\bibfnamefont {I.}~\bibnamefont {Horvath}},\ and\ \bibinfo {author} {\bibfnamefont {H.}~\bibnamefont {Neuberger}},\ }\href {https://doi.org/10.1016/S0920-5632(01)01836-9} {\bibfield  {journal} {\bibinfo  {journal} {Nucl. Phys. B Proc. Suppl.}\ }\textbf {\bibinfo {volume} {106}},\ \bibinfo {pages} {760} (\bibinfo {year} {2002})},\ \Eprint {https://arxiv.org/abs/hep-lat/0110009} {arXiv:hep-lat/0110009} \BibitemShut {NoStop}%
\bibitem [{\citenamefont {Matsui}\ \emph {et~al.}(2005)\citenamefont {Matsui}, \citenamefont {Okamoto},\ and\ \citenamefont {Fujiwara}}]{Matsui:2005uh}%
  \BibitemOpen
  \bibfield  {author} {\bibinfo {author} {\bibfnamefont {K.}~\bibnamefont {Matsui}}, \bibinfo {author} {\bibfnamefont {T.}~\bibnamefont {Okamoto}},\ and\ \bibinfo {author} {\bibfnamefont {T.}~\bibnamefont {Fujiwara}},\ }\href {https://doi.org/10.1103/PhysRevD.71.114501} {\bibfield  {journal} {\bibinfo  {journal} {Phys. Rev. D}\ }\textbf {\bibinfo {volume} {71}},\ \bibinfo {pages} {114501} (\bibinfo {year} {2005})},\ \Eprint {https://arxiv.org/abs/hep-lat/0501008} {arXiv:hep-lat/0501008} \BibitemShut {NoStop}%
\bibitem [{\citenamefont {Hayata}\ \emph {et~al.}(2023)\citenamefont {Hayata}, \citenamefont {Nakayama},\ and\ \citenamefont {Yamamoto}}]{Hayata:2023zuk}%
  \BibitemOpen
  \bibfield  {author} {\bibinfo {author} {\bibfnamefont {T.}~\bibnamefont {Hayata}}, \bibinfo {author} {\bibfnamefont {K.}~\bibnamefont {Nakayama}},\ and\ \bibinfo {author} {\bibfnamefont {A.}~\bibnamefont {Yamamoto}},\ }\href {https://doi.org/10.1103/PhysRevD.108.034511} {\bibfield  {journal} {\bibinfo  {journal} {Phys. Rev. D}\ }\textbf {\bibinfo {volume} {108}},\ \bibinfo {pages} {034511} (\bibinfo {year} {2023})},\ \Eprint {https://arxiv.org/abs/2305.18934} {arXiv:2305.18934 [hep-lat]} \BibitemShut {NoStop}%
\bibitem [{\citenamefont {Hayata}\ \emph {et~al.}(2024)\citenamefont {Hayata}, \citenamefont {Nakayama},\ and\ \citenamefont {Yamamoto}}]{Hayata:2023skf}%
  \BibitemOpen
  \bibfield  {author} {\bibinfo {author} {\bibfnamefont {T.}~\bibnamefont {Hayata}}, \bibinfo {author} {\bibfnamefont {K.}~\bibnamefont {Nakayama}},\ and\ \bibinfo {author} {\bibfnamefont {A.}~\bibnamefont {Yamamoto}},\ }\href {https://doi.org/10.1103/PhysRevD.109.034501} {\bibfield  {journal} {\bibinfo  {journal} {Phys. Rev. D}\ }\textbf {\bibinfo {volume} {109}},\ \bibinfo {pages} {034501} (\bibinfo {year} {2024})},\ \Eprint {https://arxiv.org/abs/2309.08820} {arXiv:2309.08820 [hep-lat]} \BibitemShut {NoStop}%
\bibitem [{\citenamefont {Clancy}(2024)}]{Clancy:2023kla}%
  \BibitemOpen
  \bibfield  {author} {\bibinfo {author} {\bibfnamefont {M.}~\bibnamefont {Clancy}},\ }\href {https://doi.org/10.1103/PhysRevD.110.L011502} {\bibfield  {journal} {\bibinfo  {journal} {Phys. Rev. D}\ }\textbf {\bibinfo {volume} {110}},\ \bibinfo {pages} {L011502} (\bibinfo {year} {2024})},\ \Eprint {https://arxiv.org/abs/2312.08647} {arXiv:2312.08647 [hep-lat]} \BibitemShut {NoStop}%
\bibitem [{\citenamefont {Singh}(2025)}]{Singh:2025sye}%
  \BibitemOpen
  \bibfield  {author} {\bibinfo {author} {\bibfnamefont {H.}~\bibnamefont {Singh}},\ }\Eprint {https://arxiv.org/abs/2505.20419} {arXiv:2505.20419 [hep-lat]}  (\bibinfo {year} {2025})\BibitemShut {NoStop}%
\bibitem [{\citenamefont {Chatterjee}\ \emph {et~al.}(2025)\citenamefont {Chatterjee}, \citenamefont {Pace},\ and\ \citenamefont {Shao}}]{Chatterjee:2024gje}%
  \BibitemOpen
  \bibfield  {author} {\bibinfo {author} {\bibfnamefont {A.}~\bibnamefont {Chatterjee}}, \bibinfo {author} {\bibfnamefont {S.~D.}\ \bibnamefont {Pace}},\ and\ \bibinfo {author} {\bibfnamefont {S.-H.}\ \bibnamefont {Shao}},\ }\href {https://doi.org/10.1103/PhysRevLett.134.021601} {\bibfield  {journal} {\bibinfo  {journal} {Phys. Rev. Lett.}\ }\textbf {\bibinfo {volume} {134}},\ \bibinfo {pages} {021601} (\bibinfo {year} {2025})},\ \Eprint {https://arxiv.org/abs/2409.12220} {arXiv:2409.12220 [hep-th]} \BibitemShut {NoStop}%
\bibitem [{\citenamefont {Yamaoka}(2025)}]{Yamaoka:2025sdm}%
  \BibitemOpen
  \bibfield  {author} {\bibinfo {author} {\bibfnamefont {T.}~\bibnamefont {Yamaoka}},\ }\Eprint {https://arxiv.org/abs/2504.10263} {arXiv:2504.10263 [hep-lat]}  (\bibinfo {year} {2025})\BibitemShut {NoStop}%
\bibitem [{\citenamefont {Kikukawa}\ and\ \citenamefont {Yamada}(1999{\natexlab{b}})}]{Kikukawa:1998py}%
  \BibitemOpen
  \bibfield  {author} {\bibinfo {author} {\bibfnamefont {Y.}~\bibnamefont {Kikukawa}}\ and\ \bibinfo {author} {\bibfnamefont {A.}~\bibnamefont {Yamada}},\ }\href {https://doi.org/10.1016/S0550-3213(99)00059-0} {\bibfield  {journal} {\bibinfo  {journal} {Nucl. Phys. B}\ }\textbf {\bibinfo {volume} {547}},\ \bibinfo {pages} {413} (\bibinfo {year} {1999}{\natexlab{b}})},\ \Eprint {https://arxiv.org/abs/hep-lat/9808026} {arXiv:hep-lat/9808026} \BibitemShut {NoStop}%
\bibitem [{\citenamefont {Ambjorn}\ \emph {et~al.}(1983)\citenamefont {Ambjorn}, \citenamefont {Greensite},\ and\ \citenamefont {Peterson}}]{Ambjorn:1983hp}%
  \BibitemOpen
  \bibfield  {author} {\bibinfo {author} {\bibfnamefont {J.}~\bibnamefont {Ambjorn}}, \bibinfo {author} {\bibfnamefont {J.}~\bibnamefont {Greensite}},\ and\ \bibinfo {author} {\bibfnamefont {C.}~\bibnamefont {Peterson}},\ }\href {https://doi.org/10.1016/0550-3213(83)90585-0} {\bibfield  {journal} {\bibinfo  {journal} {Nucl. Phys. B}\ }\textbf {\bibinfo {volume} {221}},\ \bibinfo {pages} {381} (\bibinfo {year} {1983})}\BibitemShut {NoStop}%
\end{thebibliography}%

\end{document}